\documentclass[aps,prl,twocolumn,showpacs,floatfix,superscriptaddress]{revtex4}
\usepackage{graphicx,subfigure}
\usepackage{amsmath,amssymb}
\usepackage{float}
\usepackage[hidelinks]{hyperref}
\usepackage{xcolor}

\usepackage{bm}

\newcommand{\be}{\begin{eqnarray}}
\newcommand{\ee}{\end{eqnarray}}

\begin{document}

\title{Correlating confinement to topological fluctuations near the crossover transition in QCD}
\author{ Rasmus N. Larsen}
\email{rasmus.n.larsen@uis.no}
\affiliation{University of Stavanger, 4021 Stavanger, Norway }
\author{Sayantan Sharma}
\email{sayantans@imsc.res.in}
\affiliation{The Institute of Mathematical Sciences, a CI of Homi Bhabha National Institute, Chennai 600113, India}
\author{ Edward  Shuryak }
\email{edward.shuryak@stonybrook.edu}
\affiliation{Physics Department, Stony Brook University, Stony Brook New York 11794,USA }

\begin{abstract}
We show the existence of strong (anti)correlations between the topological hot spots and the local values of the trace of the 
Polyakov loop in $2+1$ flavor QCD with physical quark mass, in the vicinity of the crossover transition corresponding to the 
simultaneous restoration of chiral symmetry and deconfinement. Using sophisticated lattice techniques, we have carefully 
identified the topological hot spots using quark zero modes and measured the short-distance fluctuations of the Polyakov 
loop about them, showing how the latter is repelled quite strongly around the peak of the zero modes. Though we could 
explain some aspects of these correlations within the instanton-dyon picture, our work sets the stage for a larger 
goal towards a systematic study of the role of different topological species that interact with the Polyakov loop, 
establishing the strong connection between topology and confinement.    
\end{abstract}
\pacs{2.38.Gc, 11.15.Ha, 11.15.Kc, 11.30.Rd}
\maketitle

\section{Introduction}

The two most important nonperturbative phenomena in Quantum Chromodynamics 
(QCD) are chiral symmetry breaking and confinement~
\cite{Ding:2015ona,Schmidt:2017bjt, Lombardo:2020bvn}. The corresponding 
order parameters are (the nonzero value of)  the quark condensate 
$\langle \bar q q \rangle \neq 0$ and the (vanishingly small) expectation value 
of the Polyakov loop $\langle \text{Tr} [P(\vec x)] \rangle/3 = 0$. Many authors have debated 
whether and how precisely these two phenomena can be related, see  for e.g.
~\cite{Shuryak:1982hk,Shuryak:2014gja,Larsen:2015vaa, Larsen:2015tso,
Bazavov:2016uvm,Clarke:2020htu,DeMartini:2021xkg}. In this Letter, 
we address this fundamental question \emph{at the local level}, using 
gauge configurations from lattice simulations of $2+1$ flavor QCD, by correlating 
the location of the maxima of zero modes of the Dirac operator with the local 
minima of the Polyakov loop.

Furthermore, it is now widely believed that both of these phenomena are driven by 
gauge topology~\cite{Schafer:1996wv}. In order to produce confinement, the 
topological objects need to interact with $P(\vec x)$ locally and suppress it, in the 
confined phase at $T\lesssim T_c$, where $T_c$ is the chiral crossover-transition 
temperature~\cite{Borsanyi:2010bp,Bazavov:2011nk,Bhattacharya:2014ara,Burger:2018fvb,Bazavov:2018mes}. 
With increasing temperatures, the density of these topological objects gets small, 
but still to a certain extent suppress the local fluctuations of $P(\vec x)$. 
In $3+1$ dimensions, instantons do not couple to the Polyakov loop, though it 
was historically introduced to explain confinement in $2+1$ dimensions~\cite{Polyakov:1976fu}.  
Several other candidates have been considered in literature which include 
vortices ~\cite{Nielsen:1973cs,Mandelstam:1974pi,Engelhardt:2004pf,Greensite:2016pfc},  
monopoles~\cite{tHooft:1981bkw,Polikarpov:1996wd,DiGiacomo:1999yas,Bonati:2011jv},  
and the constituents of instantons,  the so-called instanton-dyons~\cite{Kraan:1998sn,Lee:1998bb,Kraan:1998pm}. 
Well separated isolated instanton-dyons are simply three-dimensional monopoles, for which 
the gauge potential $A_4(\vec x)$ plays the role of the adjoint Higgs field. In Euclidean 
finite temperature formulation, it may have a nonzero average value, resulting in a 
nontrivial holonomy. A similar to magnetic monopoles, instanton-dyons repel the
\emph{Higgs field}  at their centers. Therefore  one  expects to see a difference 
between the average value of the Polyakov loop and its local value at the center of an 
instanton-dyon.  Similar picture exists in the core of the vortices. However, in order 
to really identify those and quantify the role which each of them plays, one can only 
rely on first principles lattice QCD calculations.

The main objective of this Letter is to unambiguously identify the correlations, if any,  
between the position of the topological zero modes of the QCD Dirac operator and the local value 
of the Polyakov loop.  We focus on a temperature range just beyond the chiral crossover transition 
where the connection between the zeromodes to a semiclassical ensemble of instanton-dyons 
has been shown previously~\cite{Gonzalez-Arroyo:1995isl, Gonzalez-Arroyo:1995ynx, 
GarciaPerez:1999hs, Bornyakov:2013iva, Larsen:2018crg, Larsen:2019sdi}.
However the observables we suggest here  and techniques which we have developed in order to 
reliably measure the correlations is very general and does not rely upon the existence of any specific 
topological ensemble. We discuss the numerical setup in the next section followed by our specific proposal 
of different observables in the subsequent section that helps one to reliably establish the sensitivity of the hot spots 
of topological fluctuations  to the local minima of the Polyakov loop.  We finally conclude with the deeper implications 
of our work, of how it impacts our understanding of the role of topology in explaining confinement in gauge 
theories.

\section{Methodology} \label{sec:method}

For success of such a study it is crucial that we use lattice  fermions with the best chiral properties on 
the lattice. This includes maintaining an exact index theorem on the lattice, by which one can identify the 
topological objects through the zero modes of the QCD Dirac operator. It is also important that the gauge 
ensembles generated using dynamical fermions also preserve the chiral properties of the latter, without 
which there could be artifacts of explicit chiral symmetry breaking. The gauge configurations used in this 
work are $2+1$ flavor QCD configurations generated by the HotQCD Collaboration~\cite{Bhattacharya:2014ara} 
using the M\"{o}bious domain wall discretization~\cite{Kaplan:1992bt,Brower:2012vk} for fermions and Iwasaki 
gauge action.  The residual chiral violation of the configurations generated using  M\"{o}bius domain wall fermions 
is tiny, of the order of $\sim 2\times 10^{-3}$. We have used \emph{overlap fermions}~\cite{Narayanan:1994gw,Neuberger:1998my} 
as the probe to measure the topological content of the sea-gauge ensembles since the domain wall fermions 
do not satisfy an exact index theorem on the lattice.  The zero modes of the valence overlap Dirac operator 
can be identified with the topological objects of the underlying gauge configurations~\cite{Berruto:2000fx} 
since it satisfies an exact index theorem, even at finite lattice  spacings~\cite{Hasenfratz:1998ri}.

The overlap Dirac operator is defined as $D = 1-\gamma_5 sign(H_W)$ where the kernel of the sign function is 
$H_W = \gamma _5 (M - a D_W)$, $D_W$ being the massless Wilson-Dirac operator~\cite{Wilson:1974sk}.  $M$ 
is the domain wall height which is chosen to be in the interval $[0,2)$ to simulate one massless quark flavor on 
the lattice. The overlap operator satisfies the Ginsparg-Wilson relation~\cite{Ginsparg:1981bj}, $\gamma _5 D^{-1} + 
D^{-1} \gamma _5 = a \gamma _5$ which is numerically implemented to a precision of $10^{-10}$.  Furthermore 
we studied the eigenspectrum of the overlap operator by varying the periodicity phase $\phi$ of the 
valence overlap Dirac fields along the temporal torus, defined as $\psi(\tau+1/T)=\rm{e}^{i\phi}\psi(\tau)$. 
We can then identify  different species of instanton-dyons with the Dirac zero modes corresponding to 
the phases $\phi$, following the procedure outlined in~\cite{Gattringer:2002wh,Larsen:2019sdi}.

We have also carefully chosen fairly large volume lattices for our study since the topological content of the 
gauge field ensembles are sensitive to finite volume effects.   The Euclidean space-time lattice has $N_s=32$ 
sites along each of spatial direction and $N_\tau=8$ sites along the temporal direction. The spatial volume 
in physical units is $\sim (4~ \text{fm})^3$, each spatial extent about four times the pion Compton wavelength. 
The quark masses are tuned to their physical values corresponding to a Goldstone pion and kaon mass of $135$ 
and $435$ MeV respectively.  We have performed the study at two different temperatures $1.1~T_c$ and $1.2~T_c$,  
where the number of independent configurations  with topological charge $|Q|=1$ are $5, 12$ respectively. 
For this specific $Q$ sector the zero modes are particularly easier to locate precisely, which is 
important for finding its correlations with the gauge observables. We have not considered the other 
$|Q| > 1$ sectors in this work since identifying the zero mode for different boundary angles is nontrivial, 
and we will address this topic separately. Moreover, the number of configurations with $|Q| > 1$ 
was small for the higher temperature ensembles and we therefore do not expect the exclusion of these 
configurations to strongly impact our results.

The pseudocritical temperature  $T_c=155(9)$ MeV for these ensembles~\cite{Bhattacharya:2014ara}.  
Moreover for these two temperatures the lattice spacings are fine enough in the range $\sim 0.13$-$0.14$~fm, 
to ensure that cutoff effects are under control.

\section{Observables}

The aim of this work is to specifically look for correlations between topological 
zero modes and confinement, and our suggested techniques are very general, 
irrespective of the specific nature of the topological objects. However we will 
show some tantalizing connections to some of  our earlier studies on instanton-dyons~
\cite{Larsen:2018crg,Larsen:2019sdi} in the next section. 

Instanton-dyons, also called instanton-monopoles are one of the attractive candidates 
which may explain confinement~\cite{Diakonov:2009jq}.  For  $SU(N_c)$ gauge theory 
at finite temperature, instanton-dyons naturally arise as the substructures of instanton 
solutions in presence of a finite holonomy, also known as Kraan-van Baal-Lu-Liu 
calorons~\cite{Kraan:1998sn,Lee:1998bb,Kraan:1998pm,Gonzalez-Arroyo:2019wpu}.
Several lattice studies have reported the presence of dyons in pure gauge theories~
\cite{GarciaPerez:1999ux,Chernodub:1999wg,Gattringer:2002wh,Bornyakov:2014esa} 
as well as in QCD~\cite{Bornyakov:2015xao,Bornyakov:2016ccq}. Using sophisticated fits 
to lattice data,  it is now possible to identify the different species of instanton-dyons 
and measure their typical separations as a function of temperature~\cite{Larsen:2019sdi}. 
Clusters of local topological fluctuations identified using a local definition of topological 
charge constructed out of the eigenvectors of the valence Dirac operator with 
generalized periodicity phases, have been observed to be correlated with the
timelike monopole currents in $SU(3)$ gauge theory, defined in the maximal 
Abelian gauge~\cite{Bornyakov:2014esa}. Furthermore scatter plots of the real 
and imaginary parts of the Polyakov loop in the center of these 
clusters~\cite{Bornyakov:2014esa} further hinted at a deeper connection 
between these topological  excitations and the Polyakov loop.

The local Polyakov loop or the holonomy is defined 
as the product over the temporal links as $P(\vec{x})=\Pi_{x_4=1}^{N_\tau} U_4(\vec{x},x_4)$ 
whose $k$-th eigenvalue, $k\in [1,N_c]$ is simply $\rm{e}^{2i\pi\mu_k(\vec{x})}$.  Measuring this observable at 
each spacetime point is quite a challenging task as it has large contributions from the ultraviolet 
fluctuations of the gauge fields. We use $10$ steps of hypercubic (HYP) smearing~\cite{Hasenfratz:2001hp} 
to remove these ultraviolet fluctuations.  With this choice of smearing we improved the signal-to-noise 
ratio sufficiently to measure its local values. We introduce an observable which measures how well  
the local Polyakov loop operator on smeared gauge ensembles correlate with the fermion 
zero mode density which is defined as,
\begin{eqnarray}\label{eq:DeltaP}
C(\Delta P,\rho)= \int d^3\vec{x}~ \rho(\vec{x})~ \frac{1}{3}\big[\text{Tr} [P(\vec{x})]-\langle \text{Tr} [P(\vec{x})] \rangle \big]~.
\end{eqnarray}

In our previous works~\cite{Larsen:2018crg,Larsen:2019sdi}, we have shown that the 
fermionic zero modes are remarkably insensitive to the ultraviolet noise due to the higher 
momentum modes of the gauge fields.  We take advantage of this fact through our choice of  
this observable.

\section{Are topological fluctuations in QCD above $T_c$ correlated to confinement? }
\label{sec:Correlation}

\begin{figure}
\includegraphics[scale=0.38, angle=-3]{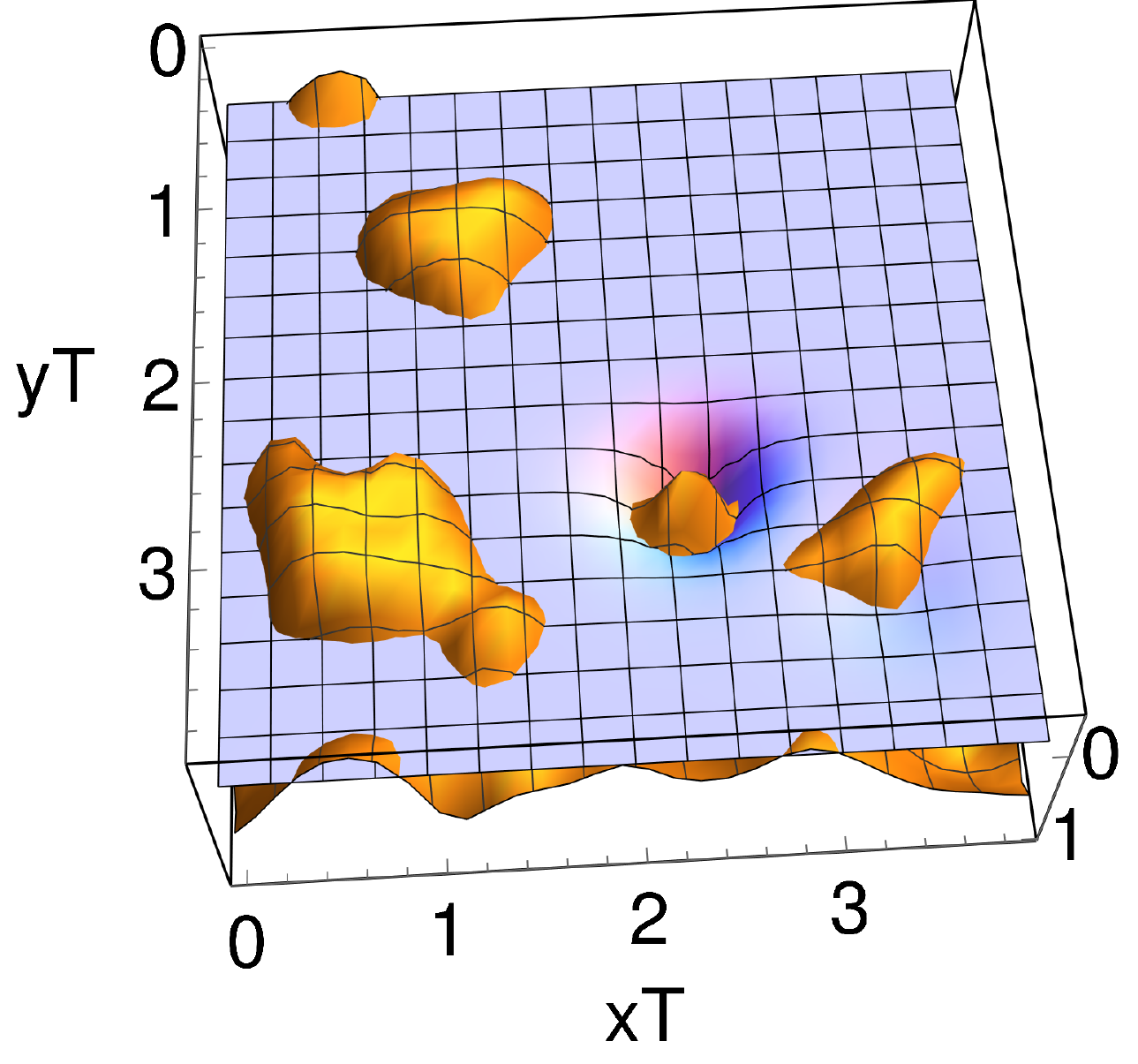}
\includegraphics[scale=0.4, angle=2]{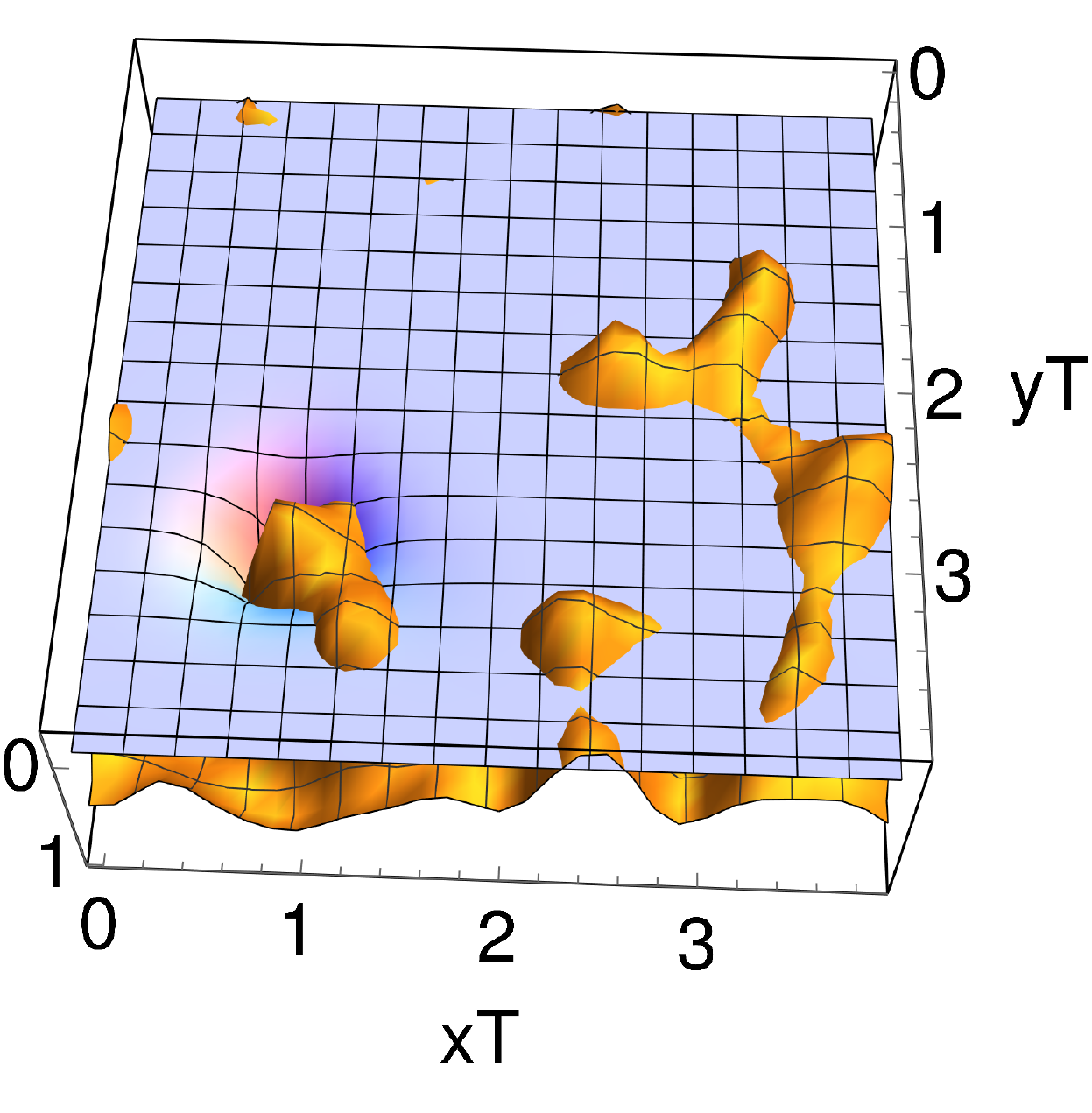}
\caption{Typical snapshots of the inverted spatial density $\rho(\vec x)$ 
of the fermion zero mode at $\phi=\pi$ (blue) superimposed on the spacetime 
profile of the inverted (real part of) Polyakov loop $\text{Tr} [P(\vec x)]/3$ (yellow), measured 
after smearing the $2+1$ flavor QCD gauge configurations at $1.2~T_c$ (above) 
and $ 1.1~T_c$ (below) respectively. The flat blue sheet corresponds 
to the zero baseline and the only zero mode is visible as a crater on this surface. 
Each plot shows that the zero mode coincides with the most localized fluctuation 
of the Polyakov loop. }
\label{fig:PloopvsFermionDensityBelow1}
\end{figure}

We first measure the local density of the fermion zero-mode wave functions and compare 
with the density profile of the trace of the Polyakov loop operator measured on smeared 
gauge ensembles. The results of such a comparison for a typical QCD configuration at 
$1.1$ and $1.2~T_c$,  for the usual antiperiodic boundary phase along the temporal 
direction, are shown in Fig.~\ref{fig:PloopvsFermionDensityBelow1}.  We have 
shown the two dimensional profiles of the quark zero modes along the $x,y$ coordinates (blue) 
superimposed over the local variation of the Polyakov loop shown in yellow. In all these plots 
the $z$ coordinate is fixed to its value at the maxima of the zero mode and the temporal direction 
is integrated out. We do indeed observe that the local Polyakov loop value drops strongly to 
negative values precisely at the location of the maxima of  the zero modes which correspond to 
$L$-dyons i.e. for the boundary phase $\phi=\pi$. This nice visual correlation is quite robust,  
and exists for all temperatures above $T_c$ we have studied.  We also see  other minima of the
local Polyakov loop; some of them might correspond to locations of other species of instanton-
dyons, but some perhaps to other topological objects e.g. vortices connecting them.  We have 
not systematically investigated their origin, leaving it for a future study.

\begin{figure}
\includegraphics[scale=0.5]{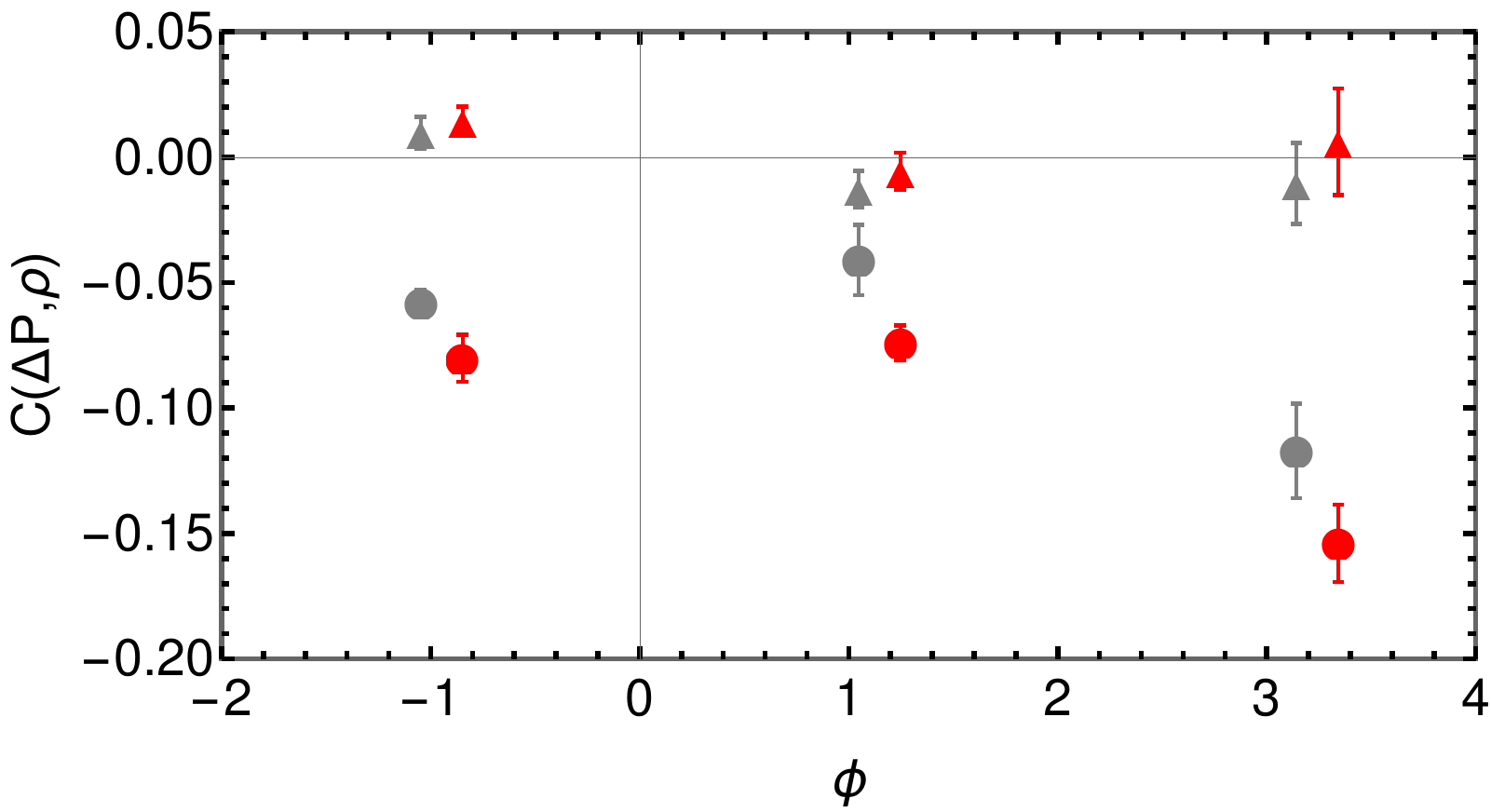}
\caption{The real and imaginary parts of $C(\Delta P,\rho)$ defined in  Eq. (\ref{eq:DeltaP}) shown as filled circles and triangles respectively for three different fermion boundary phases $\phi=\pi,\pm\pi/3$. 
The data points corresponding to  temperatures $1.1~T_c$ and $1.2~T_c$ are shown in gray and red 
respectively and the points in red are shifted  along the horizontal axis for the sake of clarity.}
\label{fig:DeltaP}
\end{figure}

Though in Fig.~\ref{fig:PloopvsFermionDensityBelow1} we have shown snapshots of correlations 
between the (real part of) Polyakov loop and topological zero modes for two configurations only, this strong 
correlation itself is very prevalent. It is present in all the configurations, at each temperature 
which we have studied so far and also for other quark periodicity phases $\phi=\pm \pi/3$. 
In order to bring out the cumulative information, we measure the observable $C(\Delta P,\rho)$ 
introduced in Eq.~(\ref{eq:DeltaP}),  which is averaged over all independent gauge ensembles 
we have examined. The typical values of the real and imaginary parts of $C(\Delta P,\rho) $ 
as a function of the fermion temporal phases are shown in  Fig.~\ref{fig:DeltaP}. Whereas its 
imaginary part stays close to zero for all three values of temporal boundary phases as expected, 
the real part is distinctly finite and negative. This again establishes the fact that local Polyakov loop values  
drop strongly at the center of the topological hot spots, but now at a cumulative level for all gauge ensembles 
at $T\gtrsim T_c$.

After demonstrating strong (anti)correlations between local hot spots of Polyakov loop with the topological 
fluctuations, we turn to some observations specific to the instanton-dyon formalism. This is motivated from 
the fact that the average values of the Polyakov loop at these temperatures have been shown previously to 
be quite accurately explained due to weakly interacting semiclassical gas of instanton-dyons~\cite{Larsen:2019sdi}.

Each constituent dyon has its own \emph{Higgsing} through a certain color projection of the local 
Polyakov loop, and we study whether such a property can be seen in the gauge configurations. 
The Polyakov loop far away  from the topological hot spots, also known as the holonomy, can be 
represented as a diagonal matrix  
$\rm{exp}[2i\pi \text{diag}(\mu_0,\mu_1,\mu_2)]$.  The constituent instanton-dyon actions 
are fractions of the  total instanton action given as $8\pi^2\nu_{m}/g^2$,  where $\nu_m=\mu_{m+1}-\mu_m$ 
represents the fractions of the circle on which the eigenvalues of the holonomy are located. For the gauge 
group $SU(3)$, there are three of them with $m= 0$-$2$ and $\mu_{3}=1+\mu_0$. 

Furthermore if instanton-dyons  are well separated, the local holonomy at the position of the 
$i$-th dyon can be written as $P(\vec{x_i})=\text{diag}\left[ \rm{e}^{i 2 \pi\mu_{i-1}}, 
\rm{e}^{i\pi(\mu_{i}+\mu_{i+1})}, \rm{e}^{i\pi(\mu_{i}+\mu_{i+1})}\right]$ ~\cite{Diakonov:2009jq}. 
In the confined phase  at $T\lesssim T_c$, where the average Polyakov loop is vanishingly small,  the instanton action is 
split evenly  between its constituent instanton-dyons,  $\nu_0=\nu_1=\nu_2=1/3$,  resulting in $\mu_0=0, \mu_1=1/3$ 
and $\mu_2=2/3$.  The Polyakov loop operator near the fermion zero mode corresponding to the $L$-dyon with 
temporal periodicity phase $\phi=\pi$ (which normalized by $2\pi$ lies between  
$\mu_1$ and $\mu_2$), is $P(\vec{x}_1)=\text{diag}[1,-1,-1]$. 
Taking a color trace and normalizing by the color factor, the imaginary part of the local Polyakov loop is  zero, 
whereas its real part is $-1/3$.  For the two M dyons which correspond to temporal boundary phases 
$\phi=\pm\pi/3$, the values of the local holonomy at its location are  $1/6 \pm i/\sqrt{12}$.

\begin{figure}
\includegraphics[scale=0.45]{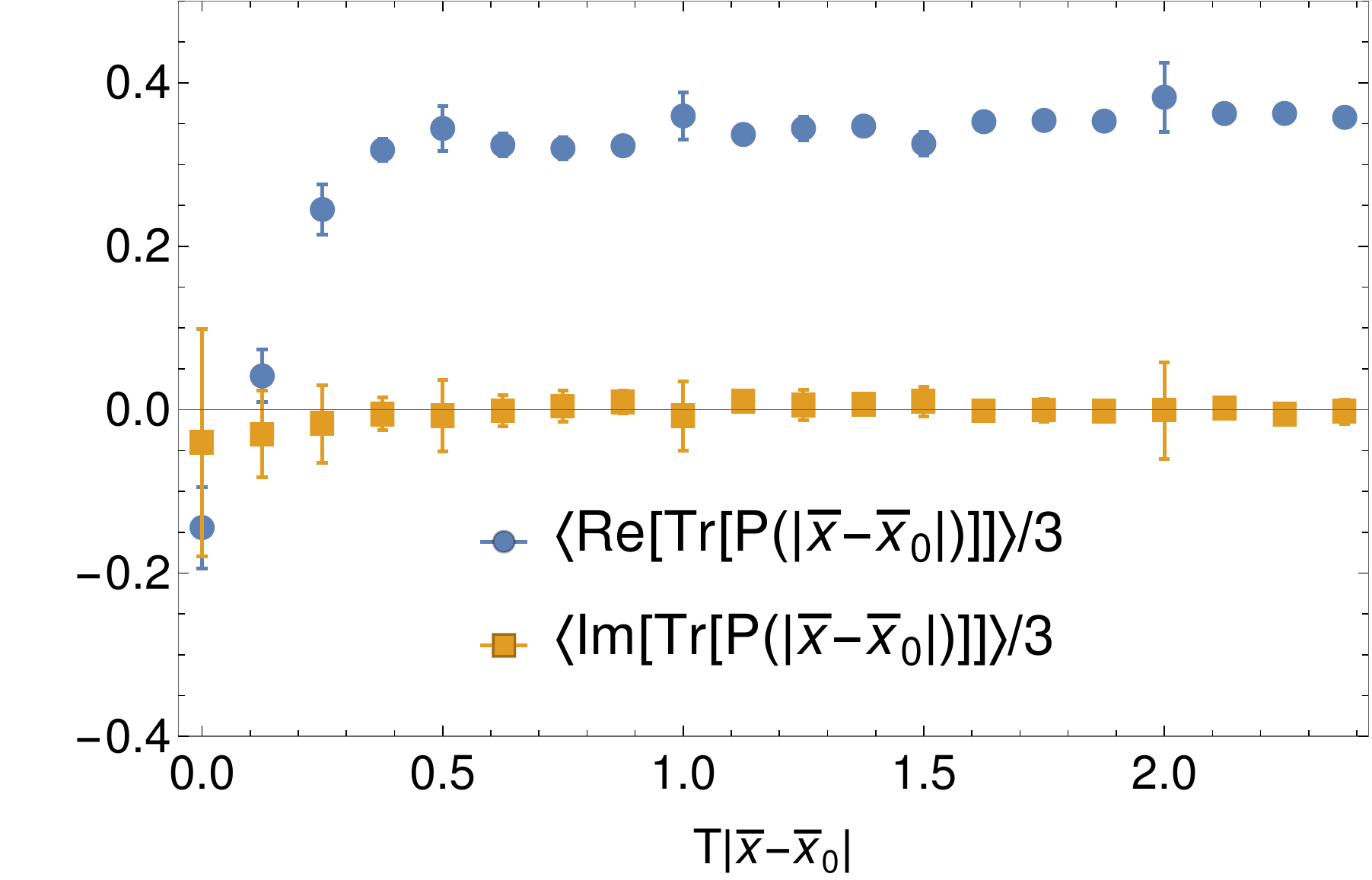}
\includegraphics[scale=0.45]{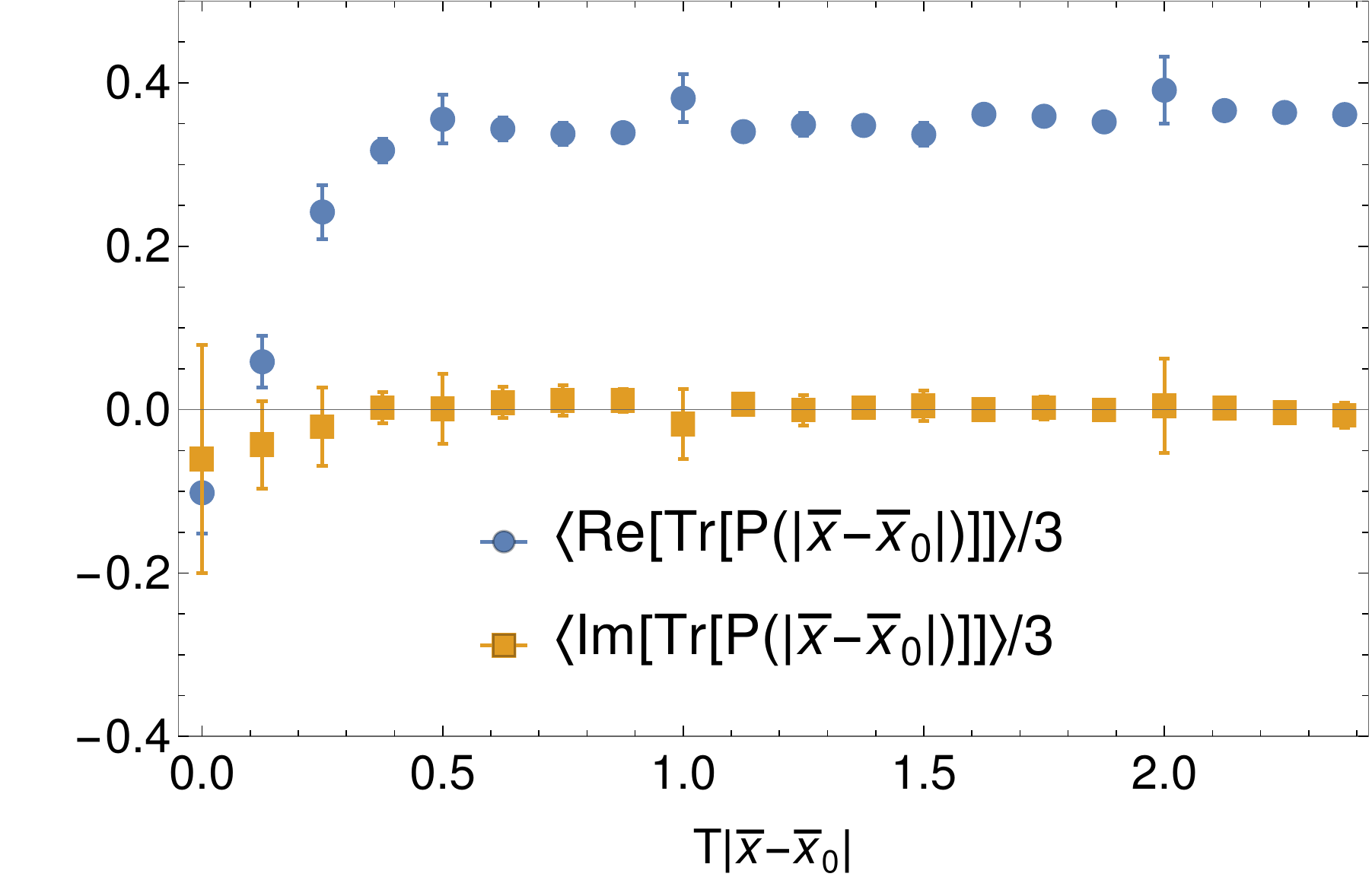}
\includegraphics[scale=0.45]{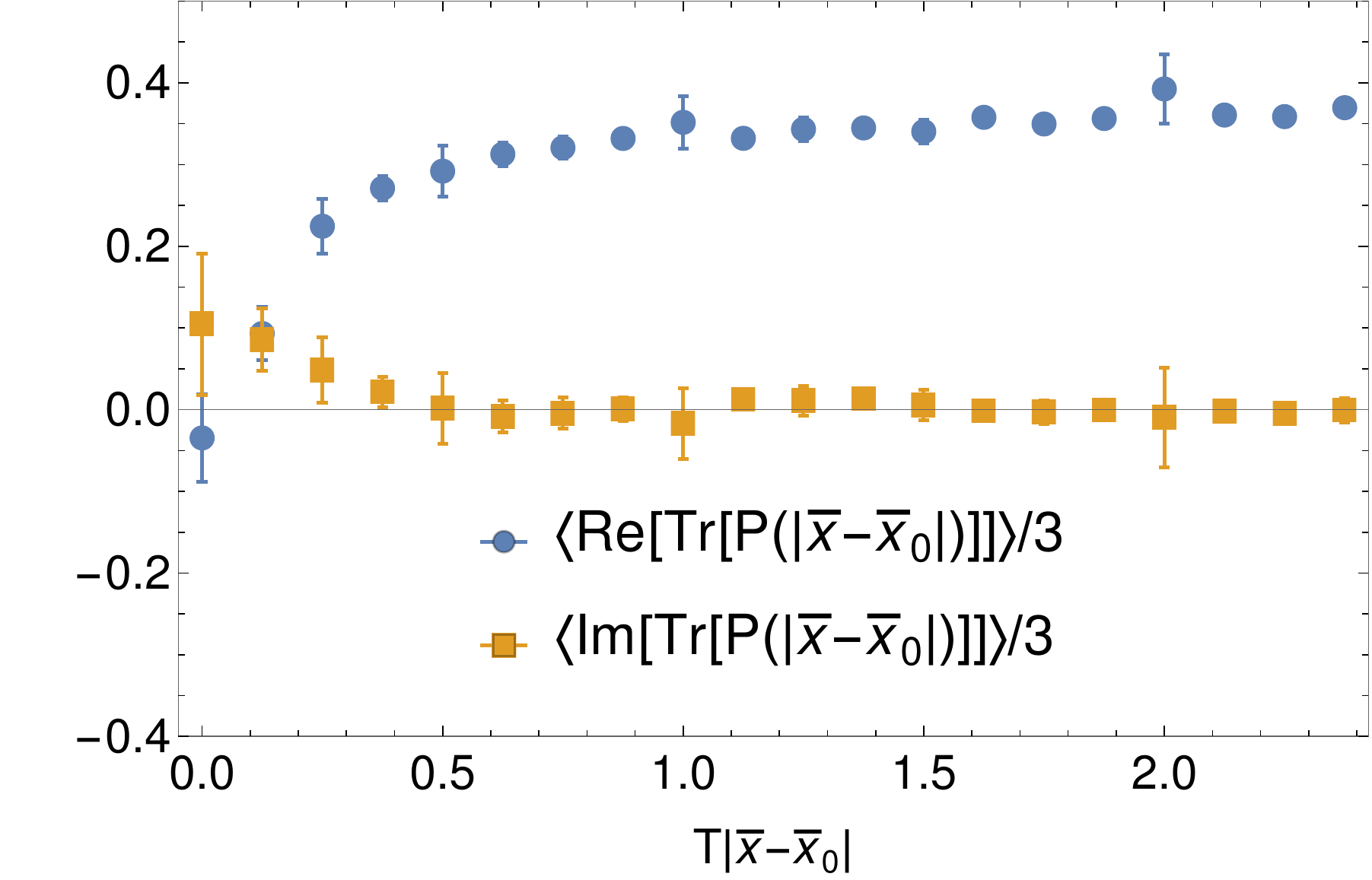}
\caption{The variation of the real and imaginary parts of the Polyakov loop $\text{Tr}[P(\vec x)]/3$ at $1.2~T_c$ for temporal 
boundary phases $\phi=\pi$ (top), $\phi=\pi/3$ (middle) and $\phi=-\pi/3$ (bottom) with the three-dimensional distance 
from the maximum of the fermionic zero modes.}
\label{fig:ReImPolyakovLoop12Tc}
\end{figure}

We next measure the real and the imaginary parts of the Polyakov loop $\text{Tr}[ P(\vec x) ]/3$ as a function of 
the distance from the center of the fermion zero modes belonging to specific species of 
instanton-dyons, for each gauge configuration and subsequently performing a statistical 
average over all the configurations studied. The results for these observables at 
$T=1.2~T_c$ are shown in Fig.~\ref{fig:ReImPolyakovLoop12Tc} for three different temporal 
phases for the valence fermions corresponding to the $L$ (top panel) and two different species of 
$M$ dyons (mid and lower panels) respectively.  For $L$ dyons,  we observe that the real part 
of the local Polyakov loop at its center (which is the origin of the plots) is negative, whereas its 
imaginary part is consistent with zero. At distances far away from the peak of the zero mode 
the real part of the local value of the Polyakov loop approaches  its average value, while the 
imaginary part remains close to zero. These results are consistent with the expectations from 
a weakly interacting instanton-dyon model close to the confining phase. 
Contrasting this with the behavior of the local Polyakov loop for temporal boundary phases 
$\phi=\pm \pi/3$, we find that its real part shows an upward trend from negative towards 
zero, respectively. The central value of the imaginary part on the other hand, changes sign 
when changing between these two boundary phases,  albeit large statistical errors. 
The data shows that the  behavior of the local Polyakov loop deviates from the expectation of 
a weakly interacting gas more so for $M$ dyons  as compared to $L$ dyons.  This can be 
due to the fact that the average Polyakov loop is not negligibly small at $1.2~T_c$ and a 
semiclassical description of instanton-dyon gas  may no longer be good.
With increased statistics, the imaginary part of the local Polyakov loop might provide a better 
understanding of  the interactions between the different species of instanton-dyons.

\section{Summary \& Outlook}

In this Letter we have shown that there exists strong (anti)correlations between the 
topological hot spots (which can be identified as instanton-dyons ) and the confinement 
order parameter $\langle \text{Tr}[P(\vec {x})]/3\rangle$ at temperatures 
$T\gtrsim T_c$. Moreover, it is a local effect. 
Revealing such correlations at a local level was made possible due to a combination of two 
novel techniques we have used. First was the efficient identification of the topological 
zero modes of the noisy $2+1$ flavor QCD ensembles. Although they were generated 
using domain-wall fermion discretization for the quarks with a relatively good chiral 
property, we identified its zero modes with valence overlap Dirac operator to make 
use of the exact index theorem for the latter.
 
Secondly we have successfully isolated the localized fluctuations of the Polyakov loop from 
the noisy large-scale fluctuations via well-tuned smearing techniques. Our study provides a 
first glimpse of how topological fluctuations due to, for e.g., instanton-dyons can result in suppressing 
the Polyakov loop values. In order to understand how topology drives confinement at a more 
quantitative level,  we would like to extend this work towards identifying the role of other 
topological objects and their interactions. Such a study will quantify from first principles,  
the origin of confinement, as  driven by gauge topology in its various forms.

\section*{Acknowledgments}

This work was supported in part by the Office of Science, U.S. Department of Energy,   
under Contract No. DE-FG-88ER40388 (E.S.). R.N.L. acknowledges support by the Research 
Council of Norway under the FRIPRO Young Research Talent grant No. 286883. We thank the 
HotQCD collaboration, formerly also consisting of members from the RBC-LLNL collaboration, 
for sharing the domain wall configurations with us. S.S. gratefully acknowledges 
support from the Department of Science and Technology, Government of India through a 
Ramanujan Fellowship. This work used the computational resources at the Institute 
of Mathematical Sciences and we thank the institute for the generous support.  Our GPU 
code is in part based on some publicly available QUDA libraries~\cite{Clark:2009wm}.

\end{document}